\documentclass[12pt]{article}

\input epsf
\usepackage{epsfig}
\usepackage{graphicx,epsf,amsmath,amsfonts,amssymb,amsbsy}

\def\i{\mbox{i}}
\def\\i{\mbox{\scriptsize{i}}}
\def\d{\mbox{d}}
\def\Im{\mbox{Im}}
\def\Re{\mbox{Re}}

\title{Resonant mode of elastic scattering of protons \\ at ultra-high
energies}
\author{M. L. Nekrasov\\
{\small\it 
Institute for High Energy Physics, NRC ``Kurchatov
Institute'',}  \vspace*{-4\baselineskip}\\
{\small\it Protvino 142281, Russia} }
\date{}
\begin{document}
\maketitle

\begin{abstract}
An interpretation is given for elastic proton scattering at ultra-high energies with impact-parameter amplitude exceeding the black disk limit. It is shown that this scattering mode can arise due to the contribution of an exceptional intermediate state that unites correlated partons of colliding protons at ultra-high energies into a single coherent system. The behavior of the real part of the amplitude in this mode is discussed.
\end{abstract}

\section{Introduction}\label{sec1}

In recent years, there has been an increased interest in diffraction
processes at ultra-high energies, see
e.g.~[1--21].
Primarily, this is due to the expectation that elastic scattering of
protons at the LHC energies is approaching pre-asymptotic. In this case,
the scattering characteristics are simplified and become largely
independent of specific details inherent in particular hadrons. On this
background, a discussion is unfolding about scenarios for the 
asymptotic behavior of the elastic proton scattering.

According to widespread belief, it corresponds to the black disk
pattern, which is characterized by complete absorption in the central
region in the impact parameter space [1--6]. Correspondingly, the elastic scattering in this region
is shadow, and real scattering is possible only at the periphery. An
approach to this mode is characterized by a monotonous ``blackening'' of
the central region and a simultaneous increase in the amplitude up to
the so-called black disk limit. Simultaneously, more sharp edge of the
central region is formed, and its transverse size grows as $\ln s$. An
important characteristic of the black disk limit is the equality of
elastic and inelastic scattering cross sections, so that
$\sigma_{e\ell}/\sigma_{tot} = 1/2$ at $s \to \infty$.

However, lately an alternative point of view is gaining more and more
weight, related to the pattern of the black ring at ultra-high energies [8--21]. The size of the ring is also
growing as $\ln s$, but inside the ring the absorption is decreasing.
Notwithstanding this the amplitude increases in this region, reaching
values exceeding the black disk limit. The reason for this behavior is
unclear. However, it is allowed by the unitarity relation and confirmed
by some data fittings at $\sqrt{s} = 7$--13 TeV [7, 16--22] (in the TOTEM analysis \cite{TOTEM16}
see results depicted in Fig.~19). A characteristic feature of this
scenario is the accelerated growth of the elastic scattering, which
leads to $\sigma_{e\ell}/\sigma_{tot}$ exceeding 1/2 at $s \to \infty$.

At the moment, both of the above scenarios are essentially equal in
terms of data description. This is explained by the ambiguity in
determining the impact-parameter amplitude based on scattering data in
view of the uncertainty in the phase of the amplitude. In this regard,
the most reliable criterion for choosing between the scenarios is the
ratio $\sigma_{e\ell}/\sigma_{tot}$. This criterion, however, only comes
into play when the edge effects are negligible and a truly asymptotic
regime is reached. The maximum energy for the cross-sections measurement
is provided by cosmic rays. The corresponding measurements at $\sqrt{s}
= 57$ TeV \cite{Cosmic} give $\sigma_{e\ell}/\sigma_{tot} = 0.31 \pm
0.19$. Unfortunately, this result is too imprecise to make a definite
conclusion, and it is not clear whether the asymptotic region
is actually reached at the specified energy.

As a result, theoretical studies come to the fore. Generally, they must
explain the above scenarios and prove the impossibility of one of them.
However, only the first scenario is clear so far. Namely, the black disk
mode occurs due to dropout of the protons from elastic scattering owing
to activation of inelastic channels in the head-on collisions. The
second scenario is still unclear. Moreover, at the moment it has no
unified appellation. In addition to the aforementioned ``black ring'' a
term ``hollowness'' is also used, which means a decrease in blackness in
the inner region compared to the ring. In the pioneering works
\cite{Troshin1993,Troshin1998}, the emphasis was placed on the anomalous
properties of the amplitude. Namely, in view of the growth of its
imaginary part simultaneously with the decrease in the absorption, the
term ``antishadow scattering'' was proposed. Then the same authors
proposed the term ``reflective scattering'' \cite{Troshin2007}, focusing
on the change by $\pi$ of the phase shift of the amplitude in the region
where it is anomalously large. In this case, an analogy was used with
the well-known in optics property of changing by $\pi$ the phase of a
reflected wave at the boundary between two different media. However, the
same property of the amplitude (the characteristic phase shift) and its
anomalously large magnitude may be associated with resonant behavior. On
this basis, the term ``resonant disk mode'' was proposed
\cite{Anisovich2014}.

In our opinion, the last term is most suitable as it reflects the
essence of the phenomenon. In this work, we reveal its physical content
precisely as a resonance effect due to the contributions of an
exceptional intermediate state. Initially, it does not have a definite
spin and mass, but effectively acquires them depending on the scattering
conditions. Moreover, this state can only arise at a sufficiently high
collision energy. The minimum energy is determined by the energy of
switching-on the mode of correlated motion of the partons in ultra-fast
protons. This mode was previously discussed in \cite{Gribov73009}.
Justification of its emergence at ultra-high energies was given in
\cite{Nekrasov1}, and its influence on the proton properties was studied
in \cite{Nekrasov2}. 

In the next section, we consider the unitarity relation and its
solutions of the type of black and resonant disks with smooth edges.
Section \ref{sec3} gives interpretation of the resonant disk solution as
a consequence of the contribution of an intermediate state. In the
final section, we discuss the results.

\section{Scattering amplitude and unitarity relation}\label{sec2}

The characteristics of the second scenario listed above are in fact
linked by the unitarity relation. At high energies it has the form
\begin{equation}\label{D1}
\Im \, h(s,b) = |h(s,b)|^2 + H_{in}(s,b) \,.
\end{equation}
Here $h(s,b)$ is the amplitude of elastic scattering in the
representation of impact parameter $b$. $H_{in}(s,b)$ is the overlap
function, which represents the sum of inelastic contributions.
The unitarity condition requires $H_{in} \ge 0$. Then it follows from
(\ref{D1}) that $H_{in} \le 1/4$, and therefore $0 \le H_{in} \le 1/4$.
Each term in (\ref{D1}) defines a profile function for the total,
elastic, and inelastic scattering, respectively. Term-by-term
calculation $4 \!\int \! \d^2 {\bf b}$ in (\ref{D1}) leads to the
identity $\sigma_{tot} = \sigma_{e\ell} + \sigma_{in}$. 

The general solution of (\ref{D1}) may be represented in the eikonal
parametrization,
\begin{equation}\label{D2}
h(s,b) = 
\frac{1}{2\i} \left[\eta(s,b) e^{2\\i\delta(s,b)} - 1\right].
\end{equation}
Here $\delta(s,b)$ is the phase shift and $\eta(s,b)$ is the parameter
of inelasticity or transparency. Both they are real, and $\eta\ge 0$.
The $\eta(s,b)$ is directly related to the overlap function,
\begin{equation}\label{D3}
\eta^2 = 1 - 4 H_{in} \,.
\end{equation} 
Hence, we have
\begin{equation}\label{D4}
0 \le \eta(s,b) \le 1 \,.
\end{equation} 
By virtue of (\ref{D2}), the imaginary and real parts of the amplitude
are
\begin{equation}\label{D5}
\Im \, h = \frac{1-\eta \cos(2\delta)}{2}\,, \quad
\Re \, h = \frac{\eta}{2} \sin (2\delta) \,.
\end{equation} 
So, $\Im \, h$ monotonically increases from $(1-\eta)/2$ to $(1+\eta)/2$
with the growth of $2\delta$ from 0 to $\pi$. 

Among the solutions of (1), a special place is occupied by those of the
disk type, which are characterized by a sharply limited scattering
region. A special role among them is assigned to purely imaginary
solutions. Proceeding from (\ref{D5}), one can specify two ways of
zeroing $\Re \, h$, the $\eta = 0$ and $2\delta = 0 \, (\mbox{mod} \,
\pi)$. In the former case, we get the so-called black disk solution:
\begin{eqnarray}\label{D6}
&\displaystyle \!\!\!\!\!\!\!\!\!\!\!\!\!\!\!
h(s,b) = \frac{\i}{2}\Theta(R-b),\quad 
H_{in}(s,b) = \frac{1}{4}\Theta(R-b), &
\\[0.4\baselineskip]\label{D7}
& \sigma_{tot} = 2 \pi R^2 \,, \quad 
\sigma_{e\ell} = \sigma_{in} = \pi R^2 \,, &
\\[0.4\baselineskip]\label{D8}
&\sigma_{e\ell}/\sigma_{tot} = 1/2\,.&
\end{eqnarray} 
In the latter case, we get solutions of the gray and resonant disks:
\begin{eqnarray}\label{D9}
&\displaystyle 
h(s,b) = \frac{\i}{2}(1 \!\mp \hat{\eta})\,\Theta(R-b), &
\\[0.5\baselineskip] \label{D10}
&\displaystyle
H_{in}(s,b) = \frac{1 \!- \hat{\eta}^2}{4}\Theta(R-b), &
\\[0.7\baselineskip] \label{D11}
& \!\!\!
\sigma_{tot} = 2 \pi (1 \!\mp\! \hat{\eta}) R^2 \,, \quad\; 
\sigma_{e\ell} = \pi (1 \!\mp\! \hat{\eta})^2 R^2 \,,
&
\\[0.7\baselineskip]\label{D12}
& \sigma_{in} = \pi (1 \!-\! \hat{\eta}^2) R^2 \,,  \quad\; 
\sigma_{e\ell}/\sigma_{tot} = (1 \! \mp \!\hat{\eta})/2 \,.&
\end{eqnarray} 
Here $\Theta(\dots)$ is the Heaviside step function, $R$ is the radius
of the disk beyond which there is no scattering ($\eta=1$, $\delta=0$).
The $\hat{\eta}$ is the transparency inside the disk. The signs $\mp$
correspond to the multiplicity of $\pi$. The ``$-$'' matches $\delta=0$
and the gray disk, the ``$+$'' does $2\delta = \pi$ and the resonant
disk. In the latter case, the $\Im \, h$ inside the disk, and also
$\sigma_{e\ell}$ and $\sigma_{tot}$, exceed those for the gray disk
despite the same $H_{in}$, and exceed those for the black disk of the
same size. These properties resemble a resonant behavior, and this
determines the name of the solution, although its nature is not yet
clear.  

Actually, the above solutions play an auxiliary role since physical
solutions must satisfy the analyticity condition. In a minimum, they should be smooth functions of $s$ and $b$. An easy way to give smoothness is to assume that $\hat{\eta}$ is a smooth function. In the case of (\ref{D9}) with the ``$-$'' sign, this allows us to direct $1\!-\!\hat{\eta}$ to zero starting from some $R_{ph}$, where $R_{ph}$ is the physical effective radius. The $R$ in this case is a technical parameter tending to infinity. (Practically, the $\Theta$-functions are discarded and $\hat{\eta}$ becomes actual transparency.) So the amplitude $h(s,b)$ and the overlap function $H_{in}(s,b)$ smoothly drop to zero outside of $R_{ph}$. The smoothed black disk is determined in this approach as the limiting case of the smoothed gray disk with $1\!-\!\hat{\eta}$ reaching 1 inside the disk.

\begin{figure*}
\hspace*{0.0\textwidth}
\includegraphics[width=0.41\textwidth]{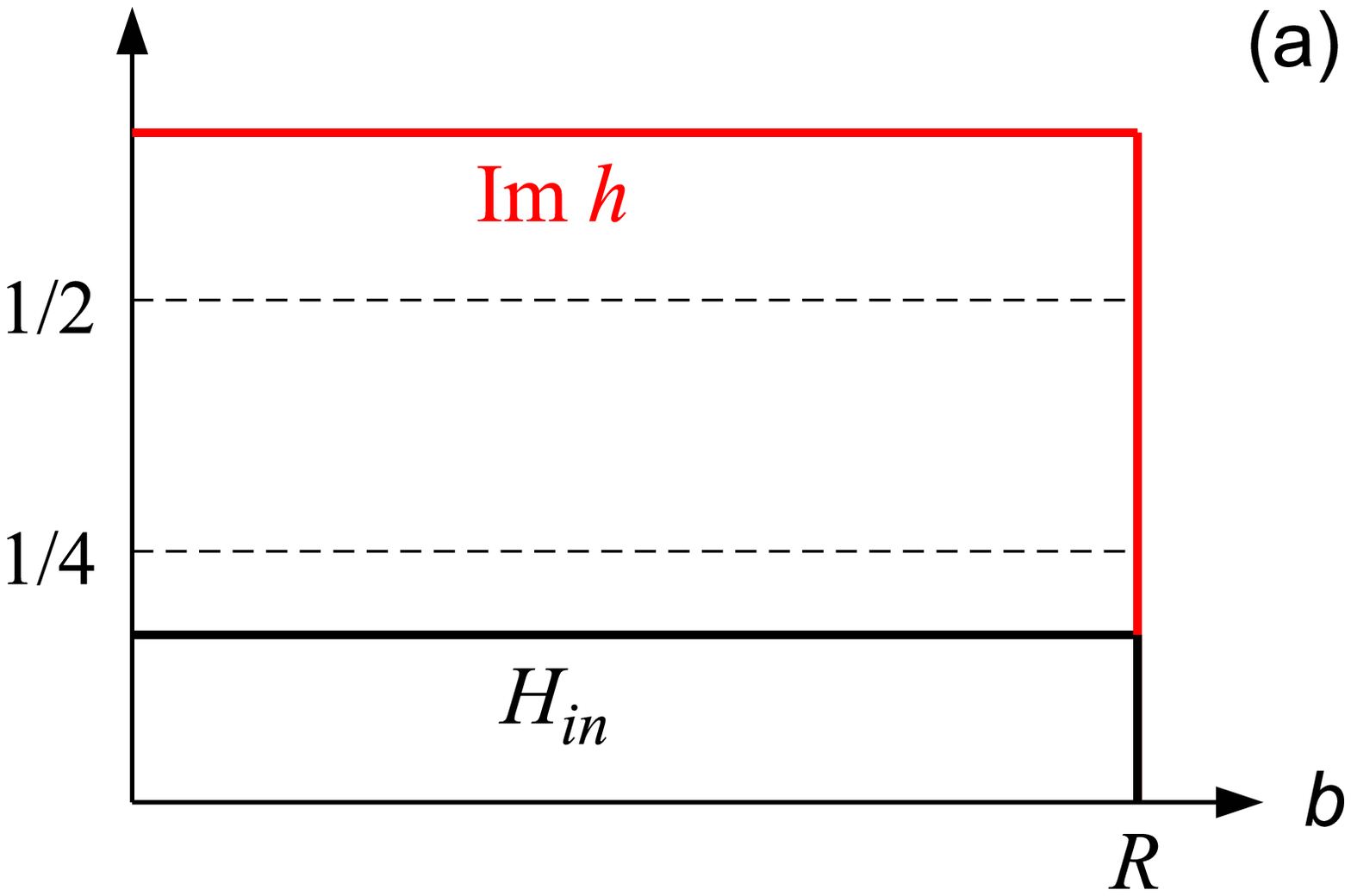}
 \hfill 
 \includegraphics[width=0.41\textwidth]{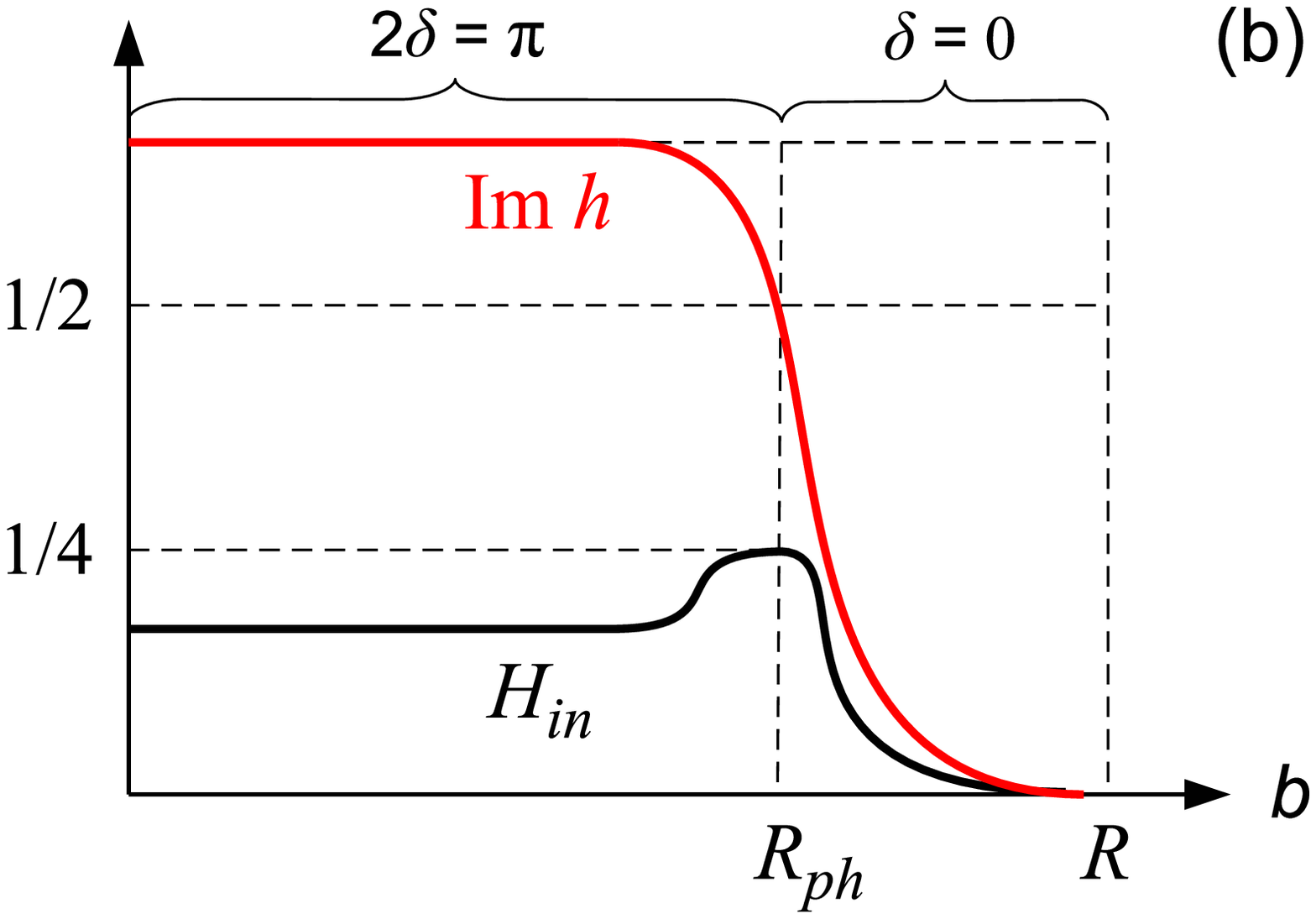}
\hspace*{0.1\textwidth}
\caption{\small The resonant disk solution of the unitarity relation
with (a) sharp and (b) smoothed edge.}
\label{Fig1} 
\end{figure*}  

In the case of the resonant disk, i.e.~solution (\ref{D9}) with the
``$+$'' sign, the above procedure cannot be directly implemented as
$1+\hat{\eta}$ cannot be reduced to zero. The problem is solved in
stages. First, we assume that when $b$ approaches $R_{ph}$ from the
inside, the $\hat{\eta}$ smoothly drops to zero, so that $1 +
\hat{\eta}$ is reduced to 1. Here $R_{ph}$ is the value of $b$ for which $\Im \, h = 1/2$. Since $\hat{\eta} = 0$, in this point a smooth transition from the $\delta = \pi/2$ mode to the $\delta = 0$ mode is possible, and we assume this occurs. With a further increase in $b$, the $\hat{\eta}$ starts to grow at the same rate as it decreased when
approaching $R_{ph}$ (this ensures the smoothness in the neighborhood of $R_{ph}$). After that, $\hat{\eta}$ can grow at any rate, providing drop of $\Im\, h$ to zero. Accordingly, the behavior of $H_{in}$ is as
follows. Inside the disk, it is less than 1/4. As $b$ approaches
$R_{ph}$, the $H_{in}$ rises to the limit of 1/4, and then decreases to
zero with a further growth of $b$. In view of the quadratic dependence
on $\hat{\eta}$, the $H_{in}$ is a smooth function of $b$ everywhere.
So, the pattern of black ring surrounding the hollowness is realized,
see illustration in Fig.~\ref{Fig1}. In fact, this behavior is a
consequence of smoothness condition. Notice that relations (\ref{D11}) and (\ref{D12}) for the cross sections cease to hold if the contributions of the edge are not negligible. 

In reality, however, the above smoothing still does not approach the
exact solution. The point is that the analyticity requires the presence
of a real part in the amplitude, exponentially decreasing at large $b$
\cite{Gribov73009}. In the case of a smoothed black disk, this can be
formally provided by adding an appropriate real function to the amplitude outside $R_{ph}$. (Simultaneously, the overlap function
$H_{in}$ must be changed outside $R_{ph}$ to provide the unitarity
condition.) In the case of resonant disk the situation is more
complicated, since the condition $\Im\, h > 1/2$ does not exclude the
presence of nonzero $\Re\, h$ inside $R_{ph}$. This can radically change the behavior of $H_{in}$. In particular, if $\Re\, h\not= 0$ at $b = R_{ph}$, then $H_{in}$ does not reach the limit of 1/4. Moreover, one can choose $\Re\, h$ such that $H_{in}$ everywhere monotonically
decreases with increasing $b$, and no hollowness appears \cite{Samokhin}. So, the condition $\Im\, h > 1/2$ does not necessary
entails the black-ring effect. Nevertheless, if $\Im\, h$ exceeds the
black disk limit, this phenomenon must be explained anyway. 

\section{Genesis of the resonant mode}\label{sec3}

Above, we noted the analogy of exceeding the black disk limit to the
resonant behavior of the amplitude. To develop the analogy, we make
appropriate transformations in formula (\ref{D2}). At first we write it
as
\begin{equation}\label{D13}
h(s,b) = \eta(s,b)
\frac{e^{2\\i\delta(s,b)} - 1}{2\i} - \frac{1-\eta(s,b)}{2\i} \,.
\end{equation}
Then we introduce the parameterization of the phase shift,
\begin{equation}\label{D14}
\tan \delta (s,b) = \frac{M \Gamma }{M^2 - s} \,.
\end{equation} 
Here $M$ and $\Gamma$ are some functions of $s$ and $b$. As a result,
the amplitude takes the form
\begin{equation}\label{D15}
h(s,b) = \frac{M \eta \Gamma}{M^2 - s -\i M \Gamma} +
\i \, \frac{1-\eta}{2} \,.
\end{equation}

If $M$ were explicitly independent of $s$ and the first term in
(\ref{D15}) would appear only at $b = (2\ell +1)/\sqrt{s}$, where $\ell$
is a positive integer, then (\ref{D15}) would represent resonance in
$\ell$ wave with mass $M$, width $\Gamma$, and the partial width $\eta
\Gamma$ in the two-proton channel. The last term in (\ref{D15}) would
mean the background contribution of non-resonant scattering.

Here it is appropriate to recall that ordinary resonances arise in the amplitude due to the effect of intermediate quasi-stationary states formed during particle collisions. The phase shift $\delta$ plays a fundamental role. Its origin is caused by the temporary capture of colliding particles by each other near the quasi-stationary state and, therefore, a delay in their passage of the interaction region \cite{Bohm,Messiah}. In this case formula (\ref{D14}) describes the behavior of $\delta$ when passing $s$ near $M^2$. The main features are the change by $\pi$ when passing through the whole resonance region, and by $\pi/2$ in the very point of the resonance. 

\begin{figure*}
\hspace*{0.0\textwidth}
\includegraphics[width=0.56\textwidth]{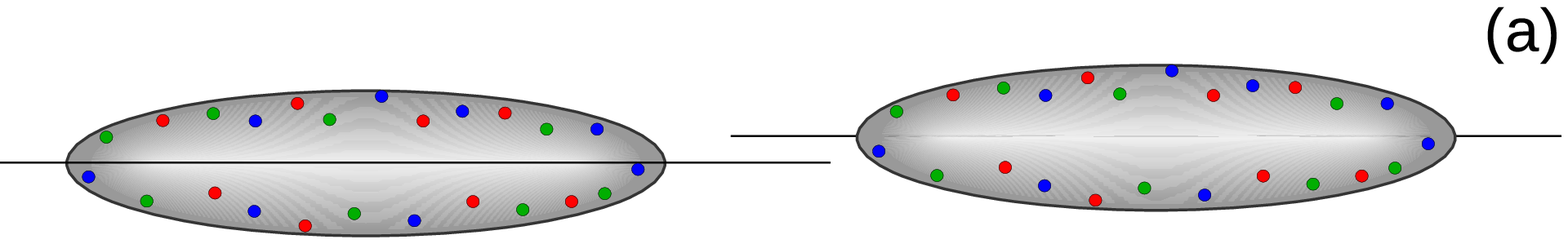}
 \hfill 
 \includegraphics[width=0.39\textwidth]{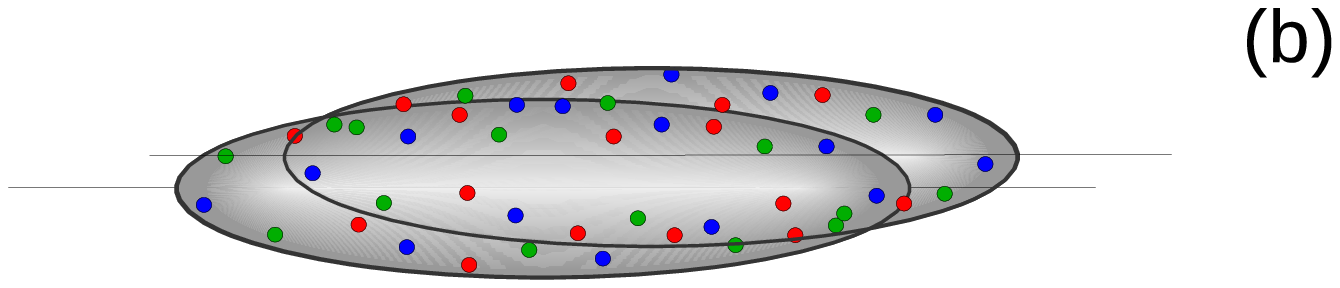}
\hspace*{0.03\textwidth}
\caption{\small (a) Each of the colliding ultra-fast protons is
represented as a rarefied cloud of very strongly coupled correlated
partons \cite{Nekrasov2}. (b) In head-on collisions, the aggregate cloud
is rarefied in the overlap region. }
\label{Fig2} 
\end{figure*}

Of course, at ultra-high energies, the ordinary resonances are not observed. However, as a hypothesis, we can assume the appearance at ultra-high energies of an exceptional intermediate state. Then, we could find the properties that it must have in order for its contribution to the amplitude steadily exceeds the black disk limit with increasing energy. As a starting point, we assume validity of formula (\ref{D14}) for the phase shift in the vicinity of resonance.

First we notice that at $M^2 = s$ the phase shift (\ref{D14}) takes the value $\delta = \pi/2$, and formula (\ref{D15}) turns into $h = \i (1\!+\!\eta)/2$. Actually, this is an amazing result since it reproduces the solution for the resonant disk. Simultaneously its structure becomes clear; it includes the contribution of the gray disk $\i (1\!-\!\eta)/2$ and an additional contribution $\i \eta$ of resonant nature.

Based on this, we conclude that our exceptional state should not have a fixed intrinsic mass, but its effective mass should be formed depending on the scattering conditions. The same should be the case with the spin of this state. Namely, the mentioned parameters must take ``running'' values $M^2 = s$ and $\ell \simeq b\sqrt{s}/2$, respectively (see below how this can be). The contribution to the amplitude of such a quasi-stationary state is resonant and regardless of specific value of the energy falls exactly at the resonance point.

In turn, the width $\Gamma$ and the partial width $\eta\Gamma$ of this state must also depend on $b$ and $s$. The exact form of this dependence is generally unknown, but some details can be deduced based on the above discussion on smoothing the resonant disk solution. Namely, since the transparency $\eta$ decreases when $b$ approaches $R_{ph}$ from the inside, the partial width must decrease, as well. (Accordingly, the fraction of the decays through inelastic channels must increase, cf.~Fig.~\ref{Fig1}). At $b = R_{ph}$, the transparency and the partial width vanish. Correspondingly, the contribution to the elastic amplitude of the above state disappears. In the case of scattering beyond $R_{ph}$, it does not appear at all, and the tail of the smoothed disc is reproduced in this region. 

Summing up, we emphasize the continuous dependence of the parameters of the above state on the scattering conditions. This prompts us to call the corresponding resonance continuous.\footnote{Note that continuous resonances are known phenomenon in nonlinear dynamics, see e.g. \cite{NonlinearDyn}.} Below we consider a model explanation of the origin and nature of this extraordinary state.

Actually, we associate its occurrence with the extra\-ordinary property of protons, which also appears with ultra-high energies. The point is that the long-known ``diffuse'' increase with the energy of the transverse sizes of protons \cite{Gribov73009,Gribov73} is accompanied by an increase in their longitudinal sizes (the quantum effect) \cite{Nekrasov2}. The latter increase is due to the permanent production of slow partons during the splitting of fast ones and the increasing in characteristic time of this process proportionally to the energy of the protons. In general, the increase in the longitudinal sizes of fast moving protons follows an increase in the longitudinal size of their interaction region [32--36]. This leads to a decrease in the volume density of the partons and, as a result, to an increase in their coupling to each other. Ultimately, this leads to the formation of a mode of correlated motion of the partons \cite{Nekrasov2}. In this mode, all the partons are relativistic and the increase in the longitudinal sizes of protons stops. Simultaneously, their expansion in the transverse directions is accelerated, changing from the diffuse law $\sqrt{\ln s}$ to $\ln s$ [24--26]. The latter property implies an increase in the growth rate of the slope of the diffraction cone from $\ln s$ to $\ln^2 \!s$. Such a change is indeed observed at the proton collision energy in the range of 2-7 TeV \cite{TOTEM19,Schegelsky-Ryskin}. Consequently, the change in the mode of the partons motion must occur in this energy range.

Below we will deal with the property of partons in the correlated motion mode, which is that they are very strongly coupled to each other and, in fact, form a rarefied cloud. In head-on collisions, the clouds overlap. However, the aggregate cloud still remains rarefied in the overlap region, see illustration in Fig.~\ref{Fig2}, and the partons of different clouds are again strongly coupled. In general, this can lead to an increase in the destruction of clouds and, accordingly, to an increase in inelastic processes with increasing energy. If no other processes occur (on the background of elastic scattering), then the black disk  pattern of scattering is realized. However, there is an alternative scenario. Namely, the correlated partons of different clouds can form a single coherent system that can exist for some time. Actually, in this case a quasi-stationary state arises, which after some time delay can again decay into two protons. The spin and mass of this state are determined by the impact parameter and the collision energy, respectively. This means appearance of a continuous resonance in the scattering amplitude.

In fact, the energy of the onset of the latter scenario must exceed the energy of the  formation of the correlated motion mode in colliding protons. The point is that the total density of the partons in the overlap region must be low enough, and therefore their coupling strong enough to form a unified coherent system. Presumably, it should approach the density of partons in each of the protons when the mode of correlated motion arises, since in this case the partons of different clouds can become mutually correlated. Fortunately, this condition may be satisfied at rather limited energies due to appearance of a hollow inside the protons during their rapid expansion in the transverse directions in the mode of correlated parton motion \cite{Nekrasov2}. (Do not confuse the hollow inside fast protons with the hollow in the overlap function.) To obtain a reliable estimate of the corresponding energy, a more detailed model for the structure of the parton clouds in fast protons is required. What is certain is that the above-mentioned energy range of 2-7 TeV for the occurrence of the correlated motion mode, can be considered as the lower bound for the energy of the onset of the resonant scattering mode.\footnote{In the phenomenological approaches the latter energy, as a rule, is not estimated (except \cite{Jenkovszky2018},  where it is $\sqrt{s} \sim 3$~TeV). Instead, the presence of resonant scattering mode at various energies, starting from 7 TeV, is indicated [7, 16--19, 21, 22].}

Another non-trivial issue is an exit from the resonant scattering mode with an increase in the impact parameter. Recall that conditions of unitarity and analyticity require that this exit be accompanied by an increase to the maximum of inelastic contributions. In our model, the exit occurs since the overlap region decreases with increasing $b$. (In accordance with the above discussion, the overlap of interior regions in protons with the reduced parton density is important.) Really, as the overlap region decreases, the influence of the partons from this region on the other partons decreases. At a certain $b = R_{ph}$ it becomes insufficient to transform the parton clouds into a single coherent system. At the same time, the aggregate coupling of the partons in the overlap region with the outer partons also decreases. So, at a certain $b$ the partons in the overlap region, having received a strong strike, can no longer be retained as part of colliding clouds. In this case the clouds are inevitably destroyed, and this process becomes dominant. The analyticity and unitarity require that this b be the same as the above $b = R_{ph}$. In this case, the central region in the impact parameter is surrounded by a ring, where the parton clouds break down. Inside the ring, due to the formation of a coherent system, the overlap function is partially converted into the resonant contribution to the amplitude.  

In fact, this picture has an independent confirmation. Namely, it was
revealed \cite{Bron2017} that the hollowness in the overlap function
must necessarily be the result of quantum coherence in underlying
processes (see also discussion in \cite{Dremin2017}). In light of this, what we have proposed is a microscopic justification for the formation of a coherent intermediate system.  
 
However, let us return to the discussion of the resonance contribution to the amplitude. Above we considered the case with a purely imaginary resulting amplitude. In the general case, a real resonant contributions may also appear if the resonant phase shift arises with a background, $\delta = \delta_{R} + \delta_{bg}$, where $\delta_{R}$ is given by (\ref{D14}). The origin of the background is associated with an additional time delay due to scattering outside the region of localization of the intermediate state \cite{Bohm}. This implies that $\delta_{bg}$ is positive and relatively~small. 

So, assuming $\delta= \delta_{R}+ \delta_{bg}$, we have
\begin{equation}\label{D16}
\frac{\eta e^{2\\i\delta} - 1}{2\i} =
e^{2\\i\delta_{bg}} \, \eta \, \frac{e^{2\\i\delta_{R}} - 1}{2\i} +
\frac{\eta e^{2\\i\delta_{bg}} - 1}{2\i} \,.
\end{equation}
Substituting (\ref{D14}) for $\delta_{R}$, we arrive at
\begin{equation}\label{D17}
h(s,b) = e^{2\\i\delta_{bg}} \frac{M \eta \Gamma}{M^2-s-\i M \Gamma} +
\frac{\eta e^{2\\i\delta_{bg}} - 1}{2\i} \,.
\end{equation}
The above formula confirms the disappearance of the resonance with vanishing $\eta$. So, the exit from the resonant mode is necessarily
accompanied by the formation of the black ring. 

Putting $M^2 = s$, we get the actual amplitude in the resonant region,
\begin{equation}\label{D18}
h(s,b) = 
\frac{\eta(s,b) e^{2\\i (\delta_{bg} + \pi/2)} - 1}{2\i}.
\end{equation}
Its imaginary and real parts are
\begin{equation}\label{D19}
\Im \, h = \frac{1+\eta \cos(2\delta_{bg})}{2}\,, \quad
\Re \, h = - \frac{\eta}{2} \sin (2\delta_{bg}) .
\end{equation} 
Since $\delta_{bg}$ is small, we have $\cos(2\delta_{bg})\!>\!0$ and,
therefore, an amplification of imaginary part of the amplitude.
Simultaneously, the real part is negative and vanishes at the boundary of the resonant region.

Immediately outside it, in accordance with the smoothness condition, the amplitude is given by formula (\ref{D2}) with $\delta = \delta_{bg}$. So the imaginary part of the amplitude is below the black disk limit, and its real part is positive. If we display the behavior of the amplitude with increasing $b$ on the complex plane, then
near $R_{ph}$ it will be represented by a curve crossing the imaginary
axis at the point $\i/2$ from the left-top to the right-down at the
angle of $2\delta_{bg}$.

\section{Discussion and conclusions}\label{sec5}

We have studied the elastic proton scattering at ultra-high energy in the mode with impact-parameter amplitude exceeding the black disk limit. This mode was previously predicted and allowed by experimental data, but its nature remained unknown.  We have explained it as resonant scattering due to the contribution of an exceptional quasi-stationary state, whose spin and effective mass are determined continuously by the impact parameter and collision energy, respectively. The appearance of such a state is possible at the energies exceeding the activation energy of the correlated motion mode of the partons in very fast protons. The latter energy, in turn, coincides with that at which the growth rate of the slope of the diffraction cone increases \cite{Nekrasov1,Nekrasov2}, which actually occurs in the region of 2-7 TeV \cite{TOTEM19,Schegelsky-Ryskin}. The partons in the mentioned mode form a rarefied cloud and are extremely strongly coupled to each other. As a result, when the clouds overlap during head-on collisions, either inelastic processes intensify, or a single coherent system of correlated partons of different clouds arises. The latter means the appearance of the quasi-stationary state mentioned above. 

In fact, both the inelastic processes and the resonant scattering can occur simultaneously with a certain probability (in the sense of quantum superposition, cf.~Eq.~(\ref{D15})). What we insist on is that the latter process can take place at ultra-high energies.  Unfortunately, it is very difficult to detect it, since the main observed effect associated with it appears at asymptotically high energies, see the discussion in Introduction. However, if the phenomenon is possible, it should be studied as fully as possible. In this article, we have taken a fundamental step in this direction and proposed a plausible explanation for its physical nature.

We hope that further development of this approach, which would include a quantitative description of the overlap of the parton clouds, could be the basis for a numerical description of the scattering amplitude and help determine which scenario is actually realized in the asymptotic energy region, black or resonant disk. At this stage, our analysis shows that if the resonant scattering scenario is confirmed, this will be evidence for the emergence of an exceptional quasi-stationary state in proton collisions, which manifests itself in the scattering amplitude as a continuous resonance.

\end{document}